\documentstyle[preprint,aps]{revtex}
\begin{document}
\preprint{\begin{minipage}{2in}\begin{flushright} SLAC-PUB-6500 (T)
 \\[-3mm] hep-lat/9405020  \end{flushright} \end{minipage}}
\input epsf
\epsfverbosetrue
\draft
\title{CORE -- A New Computational Technique for Lattice Systems}
\author{Colin~J.~Morningstar}
\address{Department of Physics \& Astronomy, University of Edinburgh,
  Edinburgh EH9 3JZ, Scotland}
\author{Marvin Weinstein}
\address{Stanford Linear Accelerator Center, Stanford University,
  Stanford, California 94309}
\date{\today}
\maketitle
\begin{abstract}
The COntractor REnormalization group (CORE) method, a new approach to
solving Hamiltonian lattice systems, is introduced.  The method
combines contraction and variational techniques with the real-space
renormalization group approach.  It applies to lattice systems of
infinite extent and is ideal for studying phase structure and
critical phenomena.  The CORE approximation is systematically
improvable and can treat systems with dynamical fermions.
The method is tested using the 1+1-dimensional Ising model.
\end{abstract}
\narrowtext

Many problems in particle and condensed matter physics cannot be
studied with conventional perturbation theory.  Aside from Monte Carlo
simulations, few tools allow one to deal with general Hamiltonian
systems and fewer tools deal directly with their infinite-volume behavior.
This paper introduces a new tool, the COntractor REnormalization
group (CORE) approximation, which can handle this class of problems.
The CORE approach is a simple, systematic procedure for improving
any Hamiltonian real-space renormalization group calculation.  Its
virtues are: it is a variational procedure which is systematically
improvable to any desired degree of accuracy; it applies to lattice systems
of infinite extent, allowing direct study of phase structure and critical
phenomena; it provides tools for error estimation; it requires modest
computer resources by modern standards; it is complementary to
Monte Carlo methods; systems with dynamical fermions can be treated.

We start with a brief description of the CORE approximation, then
illustrate and test the method in two different applications to the
1+1-dimensional Ising model.

\paragraph*{Basic Ideas}

Choosing a good trial state is crucial to the success of any variational
calculation, especially one involving a large number of degrees of
freedom.  The Hamiltonian real-space renormalization group (RSRG)
method \cite{rsprengroup} is an algorithm for constructing a class of
trial states appropriate for lattice systems.  In this approach, one
partitions the lattice into blocks containing a few sites and diagonalizes
the Hamiltonian associated with each block.  One then {\em thins} the
Hilbert space by discarding all states except those which are tensor
products of some chosen subset of low-lying block eigenstates, and an
effective Hamiltonian which describes the mixing of the retained states
is computed.  This truncation process is iterated until the effective
Hamiltonian evolves into a fixed form which can be easily diagonalized.

Unfortunately, simple RSRG truncation procedures tend to severely
underestimate the effects of block-to-block couplings and this hinders
the accurate description of long-wavelength modes on the full lattice.
Past approaches to overcoming this problem have concentrated on using
larger blocks, increasing the number of states retained per block,
or introducing more sophisticated truncation schemes
\cite{rgimprove,isingprev}.
The $t$-expansion has also been used \cite{texpan}.  The CORE
approximation is a new approach to this problem which emphasizes
simplicity, versatility, and insensitivity to the precise details of
the truncation scheme.  This insensitivity, or robustness, frees one
from the need to develop clever truncation algorithms or to retain
many states per block.  This feature of CORE, combined with its simplicity,
greatly enhances its usefulness for higher dimensional systems.
CORE also allows the use of manifestly gauge-invariant RSRG schemes
when studying lattice gauge theories; such simple schemes cannot be
exploited in the naive multi-state approach since gauge non-invariant
states are necessary for coupling neighboring blocks after the first
truncation step.  Furthermore, the CORE method does not suffer from the
series reconstruction difficulties which plague the $t$-expansion.

The basic idea of the CORE approach is to use contraction techniques
to steer the RSRG iteration.
In the limit $t \rightarrow \infty$, the operator $e^{-tH}$
{\it contracts\/} any trial state $\vert\Phi_{\rm var}\rangle$ onto
the lowest eigenstate of $H$ with which it has a non-vanishing overlap.
Therefore, the expectation value
\begin{equation}
 {\cal E}(t) = \frac{\langle\Phi_{\rm var}\ \vert e^{-tH} H e^{-tH}\vert
   \ \Phi_{\rm var} \rangle}{
   \langle \Phi_{\rm var}\ \vert\ e^{-2tH}\ \vert\ \Phi_{\rm var} \rangle}
\label{eoft}
\end{equation}
tends to the corresponding eigenvalue $\epsilon_0$ of $H$ as $t$
becomes large.  In general, ${\cal E}(t)$ cannot be computed exactly.
Reliably approximating ${\cal E}(t)$ is an integral part of the steering
process in the CORE method.

An important step in building a CORE approximation to ${\cal E}(t)$
is to construct an easily {\em computable} operator $T(t)$ which closely
approximates $e^{-tH}$ for $t$ in some range $ 0 < t < t_{max}$.
To find such an operator \cite{symapprox}, first divide
$H$ into two (or more) parts, {\it i.e.}, $H = H_1 + H_2 $,
where the individual parts $H_1$ and $H_2$ are chosen such that
$e^{-tH_1}$ and $e^{-tH_2}$ can be computed exactly.  Next,
rewrite $e^{-tH}$ as a symmetric product
\begin{equation}
 e^{-t H} = e^{-tH_1/2}\,e^{-tH_2/2}\,e^{C_3(t)}\,e^{-tH_2/2}\,e^{-tH_1/2},
\label{tzerooft}
\end{equation}
where $C_3(t)$ is a sum of terms all of which begin in order $t^3$ or
higher.   The simplest $T(t)$ is obtained by replacing $e^{C_3(t)}$ by
the identity operator.  One way to construct a better approximation is
to retain low-order terms in $C_3(t)$ and rewrite the exponential of
these operators as a symmetric product of explicitly computable terms.
Another is to use the operators $ T_p(t) = \left[\, T(t/p)
\,\right]^p$.  In any case, it is very important to ensure the
approximate contractor satisfies all the symmetries of $H$.

Given a contractor $T(t)$, $\epsilon_0$  can then be
bounded from above by computing
\begin{equation}
  {\cal E}_T(t) = \frac{\left\langle \Phi_{\rm var} \left| \ T(t)\,H\,T(t)\
   \right| \Phi_{\rm var} \right\rangle}{\left\langle \Phi_{\rm var} \left|
  \ T(t)^2\ \right| \Phi_{\rm var} \right\rangle}.
\label{eoftapprox}
\end{equation}
A best estimate for $\epsilon_0$ is obtained by minimizing
${\cal E}_T(t)$ with respect to $t$ and any parameters in
$\vert\Phi_{\rm var}\rangle$.  For a trial state
$\vert\Phi_{\rm var}\rangle=\sum_{j=1}^n \alpha_j \vert\phi_j\rangle$,
where $\{\vert\phi_j\rangle\}$ is some set of orthonormal states, one
can show that minimizing ${\cal E}_T(t)$ with respect to
the $\alpha_j$ parameters is equivalent to solving the generalized
eigenvalue problem
\begin{equation}
 \det\left( [ \! [ T(t) H T(t)] \! ] - \lambda [ \! [ T(t)^2] \! ]
 \right) = 0,
\end{equation}
where $[ \! [\dots] \! ]$ denotes truncation to the subspace spanned by
the $\vert\phi_j\rangle$ states.  In particular, for an operator $O$,
$[ \! [ O ] \! ] = POP^\dagger$ where $P$ is the projection operator
$P=\sum_{j=1}^n \vert\phi_j\rangle\langle\phi_j\vert$.  Thus, finding
the best trial state $\vert\Phi_{\rm var}\rangle$ is equivalent to
diagonalizing the {\em effective Hamiltonian}
\begin{equation}
 H_{\rm eff}(t) = [ \! [\, T(t)^2 \,] \! ]^{-1/2}\
      [ \! [\, T(t)\,H\,T(t) \,] \! ]
      \ [ \! [\, T(t)^2 \,] \! ]^{-1/2} .
\label{newham}
\end{equation}
Developing {\em this} operator in the RSRG iteration instead of
$[ \! [\, H \,] \! ]$ is a key innovation of the CORE approach.

The effective Hamiltonian defined by Eq.~\ref{newham} cannot
be exactly computed.  Another novel feature of the CORE approach
is the use of the finite {\it cluster} method to evaluate $H_{\rm eff}(t)$.
In this method, $H_{\rm eff}(t)$ (or any other {\it extensive} quantity)
is calculated as a sum of finite-volume contributions
(see Ref.~\cite{colinspapers}).  The finite cluster method, which
will be described later when applying CORE to the Ising model,
is simple to implement, provides numerous computational checks, and
does little or no harm to the variational bound in ${\cal E}_T(t)$.

The final ingredient in the CORE approximation is the selection of
a best value for $t$ in each RSRG step.  This can be done in a number
of ways.  One can extract the coefficient of the identity
operator in $H_{\rm eff}$ and vary $t$ to minimize this quantity.  Better
yet, one can evaluate $H_{\rm eff}$ in a simple product state to produce
a mean-field estimate of the ground state energy and minimize this
with respect to $t$.

In summary, the CORE method generates a sequence of effective Hamiltonians
$H_{\rm eff}^{(n)}(t_n^\ast)$ by successive thinning of degrees of freedom
using the recursion relation
\begin{equation}
 H_{\rm eff}^{(n+1)}(t)
 = R_n(t) [ \! [ T^{(n)}(t) H_{\rm eff}^{(n)}(t^\ast_n)
   T^{(n)}(t) \,] \! ] R_n(t),
 \label{newhameff}
\end{equation}
where $R_n(t)= [ \! [\, T^{(n)}(t)^2 \,] \! ]^{-1/2}$ and
the contractor $T^{(n)}(t)$ is constructed to approximate
$\exp[-tH_{\rm eff}^{(n)}(t_n^\ast)]$.  Eq.~\ref{newhameff} is evaluated
using the finite cluster method and a best $t=t_{n+1}^\ast$ must be chosen.
As the recursion proceeds, the effective Hamiltonian evolves eventually
into a simple form which can be trivially diagonalized, yielding
estimates of the ground state energy and the energies of some low-lying
excited states.

The expectation value of an extensive operator $O$ can also be
evaluated in the CORE method.  One develops $O$ using the same RSRG
transformations as for $H$, producing a sequence of effective operators
$O_{\rm eff}^{(n)}(t^\ast_n)$.  The matrix element of $O_{\rm eff}$
is then evaluated once $H_{\rm eff}$ has evolved to the point where
its ground state can be easily determined.

Note that from a programming point of view, CORE calculations involve
mainly matrix multiplications; diagonalizations and inversions of only
very small matrices are required.  Often the matrices will be sparse
and one can exploit efficient algorithms for multiplying them.

\paragraph*{The 1+1-Dimensional Ising Model}

We illustrate and test the CORE approximation in two different
applications to the 1+1-dimensional Ising model.  The Hamiltonian
in this model is given by
\begin{equation}
   H_{\rm Ising} = -\sum_j\left[ c_\lambda\sigma_z(j)
   + s_\lambda \sigma_x(j)\sigma_x(j+1)\right],
   \label{IsingHam}
\end{equation}
where $j$ labels the sites in the infinitely long chain,
$c_\lambda\!=\!\cos(\lambda \pi/2)$, and $s_\lambda\!=\!\sin(\lambda\pi/2)$,
for $0\!\leq\!\lambda\!\leq\!1$.  This model exhibits a second-order
phase transition at $\lambda\!=\!1/2$.  For $\lambda\! <\! 1/2$, the ground
state is unique and the order parameter
$\langle\sigma_x(j)\rangle\!=\!0$, for some site $j$.
When $\lambda\! > \! 1/2$, the ground state
is twofold-degenerate and the order parameter takes values
$\langle\sigma_x(j)\rangle=\pm(1-\cot^2(\lambda\pi/2))^{1/8}$.

The CORE approximation is best applied in the following sequence
of steps: (1) choose an RSRG algorithm by specifying how to partition
the lattice into blocks and which states to retain on each block;
(2) specify the truncation order in the cluster expansion of
$H_{\rm eff}$; (3) deduce the general form of $H_{\rm eff}$ based
on the choices made in steps 1 and 2;
(4) construct a contractor $T(t)$ which closely approximates
$\exp(-tH_{\rm eff})$ and is easily computable; (5) choose a
method of determining the optimal value of $t$ in each RSRG step;
(6) iteratively compute $H_{\rm eff}$ using
Eq.~\ref{newhameff} with initial condition $H_{\rm eff}^{(0)}=H$,
where $H$ is the Hamiltonian of interest, until $H_{\rm eff}$ can
be easily diagonalized.

In our first application, we partition the lattice into two-site
blocks and truncate the Hilbert space to the lowest two eigenstates
in each block.  Since our intention here is to carry out only the
{\em simplest} of calculations, we choose to truncate the cluster
expansion of $H_{\rm eff}(t)$ after three-block clusters.
The general form of $H_{\rm eff}(t)$ may then be deduced by
considering how it is computed in the finite cluster method.

Evaluation of $H_{\rm eff}(t)$ by the finite cluster method is
accomplished in the following sequence of steps.  First, compute
$H_{\rm eff}(t)$ using Eq.~\ref{newham} on a sub-lattice which contains
only a single block. This yields
\begin{equation}
 h_{\rm eff}^{(1)}(t) = H_{\rm eff}^{(1)}(t) =
 c^{(1)}_u(t)\ u + c^{(1)}_z(t)\ \sigma_z,
\end{equation}
where $u$ is a $2\!\times\! 2$ identity matrix.
Next, calculate $H_{\rm eff}(t)$ for a theory defined on a sub-lattice
made up of two adjacent blocks.  This Hamiltonian
takes the form
\begin{eqnarray}
 H_{\rm eff}^{(2)}(t) &=& c^{(2)}_u(t)u^Lu^R \!+\! c^{(2)}_z(t)
  (\sigma^L_zu^R \!+\! u^L\sigma^R_z)\nonumber\\
  &+&\! c^{(2)}_{zz}(t)\sigma^L_z\sigma^R_z
  \!+\! c^{(2)}_{xx}(t)\sigma^L_x\sigma^R_x
  \!+\! c^{(2)}_{yy}(t)\sigma^L_y\sigma^R_y,
\end{eqnarray}
where $L$ and $R$ refer to the left and right blocks, respectively,
in the cluster.  Remove from the two-block calculation those
contributions which arise from terms already included in the
single-block calculation:
\begin{equation}
 h_{\rm eff}^{(2)}(t) = H_{\rm eff}^{(2)}(t) \!-\! u^L\!\otimes
  h_{\rm eff}^{(1R)}(t)\!-\!h_{\rm eff}^{(1L)}(t)\otimes u^R.
\end{equation}
Repeat this procedure for sub-lattices containing successively more
connected blocks, then sum the contributions $h_{\rm eff}^{(m)}(t)$
from these sub-lattices with weights given by the number of ways each
sub-lattice can be embedded in the full lattice.
The stage at which one cuts off this cluster expansion
determines the maximum range of the interactions which will appear in
$H_{\rm eff}$.  For our choices, the effective Hamiltonian in this
model takes the general form
\begin{eqnarray}
  H_{\rm eff}(t) &=& - \sum_{\alpha,j} c_\alpha(t) O_\alpha(j),
 \label{Heff}\\
  O_\alpha(j) &=& \sigma_{\alpha_0}(j)\sigma_{\alpha_1}(j+1)\dots
         \sigma_{\alpha_r}(j+r), \label{Heffb}
\end{eqnarray}
where $c_\alpha(t)$ are the couplings, $\alpha$ labels the different
types of operators, and $j$ is a site label.  There are
only two one-site operators:  $\alpha^{(1)}=\lbrace u,z\rbrace$,
where $u$ denotes the identity operator.  In other words, the only
one-site operators are $O_u(j)=\sigma_u(j)=1$ and $O_z(j)=\sigma_z(j)$.
There are three two-site operators: $\alpha^{(2)}=\lbrace xx,yy,
zz\rbrace$.  The three-site operators are $\alpha^{(3)}=\lbrace
xzx,xux,xxz,zxx,yzy,yuy,yyz,zyy,zuz,zzz\rbrace$.

Our first contractor is built using the approximation
$\exp\left[-tH_{\rm eff}(t^\ast)\right] \approx S^\dagger(t) S(t)$,
where the operator $ S(t) = \prod_\alpha \{\prod_j \exp[t c_\alpha(t^\ast)
O_\alpha(j)/2] \}$.  The operators in the $\alpha$ product are ordered
according to their site range, increasing in size from right to left.
This operator can be simplified using
$\exp[y O_\alpha(j)]=\cosh^N\! y\ [1+\tanh y\ O_\alpha(j)]$,
where $N$ is the (infinite) number of sites in the lattice.
Discarding the unimportant $\cosh^N\! y$ factors, one obtains
a contractor given by $T_1(t)=S_1^\dagger(t) S_1(t)$,
where $S_1(t)=\prod_\alpha \{ \prod_j \left[ 1 + \tanh(c_\alpha t/2)
O_\alpha(j) \right]\}$.

Lastly, $t$ is chosen in each RSRG step to minimize the expectation
value of $H_{\rm eff}(t)$ evaluated in the mean-field state given by
$\vert\psi_{\rm mf}\rangle = \prod_j (\cos\theta \vert\!\uparrow_j\rangle
+e^{i\phi}\sin\theta\vert\!\downarrow_j\rangle)$, where
$\sigma_z(j)\vert\!\uparrow_j\rangle=\vert\!\uparrow_j\rangle$ and
$\sigma_z(j)\vert\!\downarrow_j\rangle=-\vert\!\downarrow_j\rangle$.
The matrix element $\langle\psi_{\rm mf}\vert H_{\rm eff}(t) \vert
\psi_{\rm mf}\rangle$ is minimized with respect to
$t$, $\theta$, and $\phi$, simultaneously.

For our second application of the CORE method, the lattice is
divided into blocks containing three sites and the Hilbert
space is again truncated to the lowest two eigenstates in each block.
The cluster expansion is taken only to three-block
clusters, so $H_{\rm eff}$ takes the general form shown in
Eqs.~\ref{Heff} and \ref{Heffb}.  We use an approximate contractor
given by $T_2(t)=S_2^\dagger(t) S_2(t)$
with $S_2(t)=\exp(-tV/2)\exp(-tH_b/2)$, where $H_b$ contains all intra-block
interactions and $V$ contains all inter-block operators (those which
cross block boundaries).  Note that $\exp(-t H_b/2) = \prod_{p}
\exp(-t H_b(p)/2)$ and $\exp(-t V/2) = \prod_{p} \exp(-t V(p)/2)$,
where $H_b(p)$ contains all operators which solely act on block $p$
and $V(p)$ contains only interactions
between block $p$ and $p\!+\! 1$.  The operators $H_b(p)$ and $V(p)$
can be exponentiated numerically with no difficulty.
We fix $t$ by minimizing the expectation value of $H_{\rm eff}$
in the mean-field state $\vert\psi_{\rm mf}\rangle$ as described
previously.

\begin{figure}
\begin{center}\leavevmode\epsfbox[80 300 530 760]{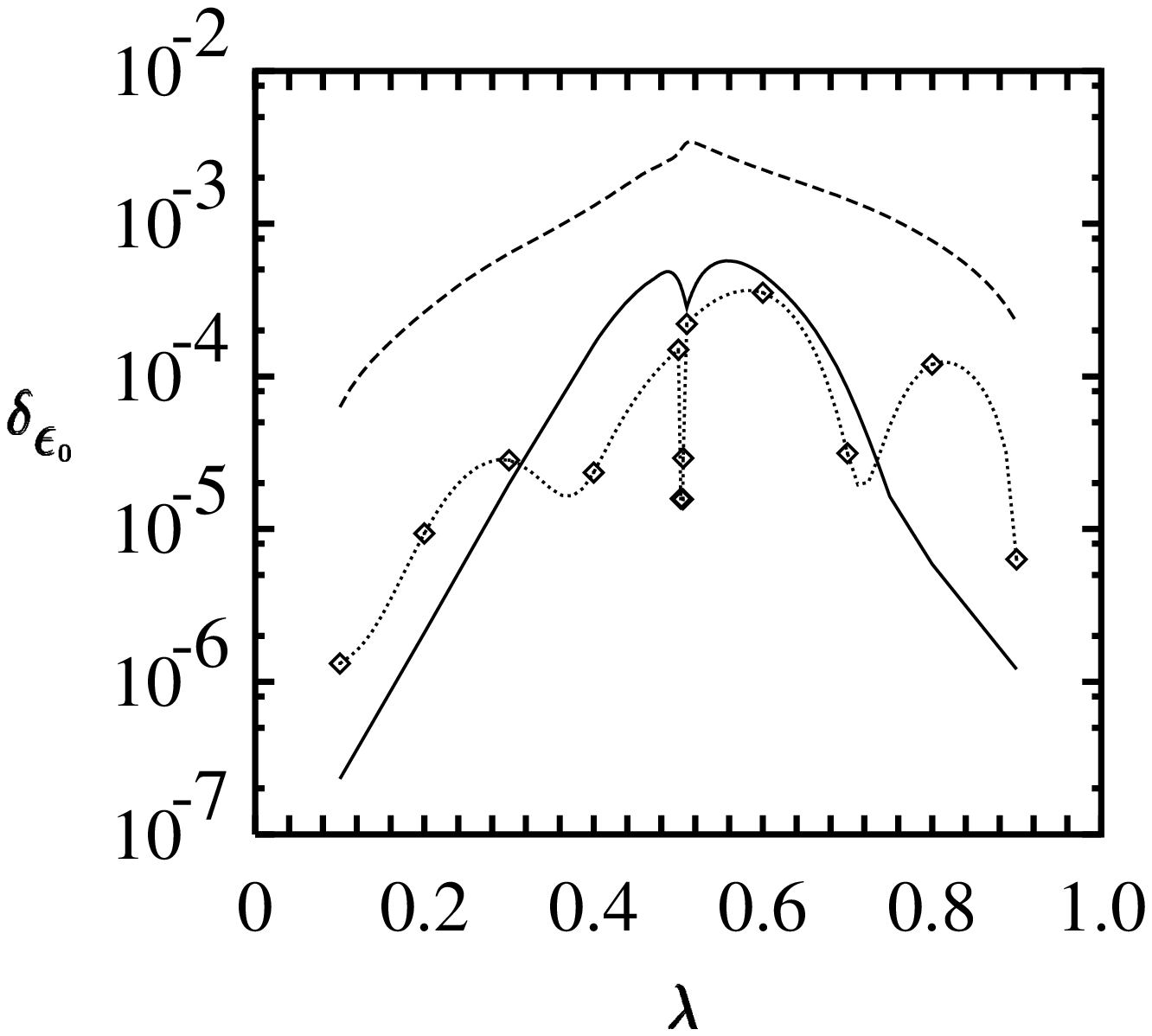}\end{center}
\caption[figenergies]{Fractional error $\delta_{\epsilon_0}$ in the
ground-state energy density estimates against $\lambda$.  Results
using $T_1^2$ (dashed curve), $T_1^{16}$ (solid), and $T_2^{12}$
(diamonds with dotted curve, to guide the eye) are shown.}
\label{figenergies}
\end{figure}

\begin{figure}
\begin{center}\leavevmode\epsfbox[80 300 530 760]{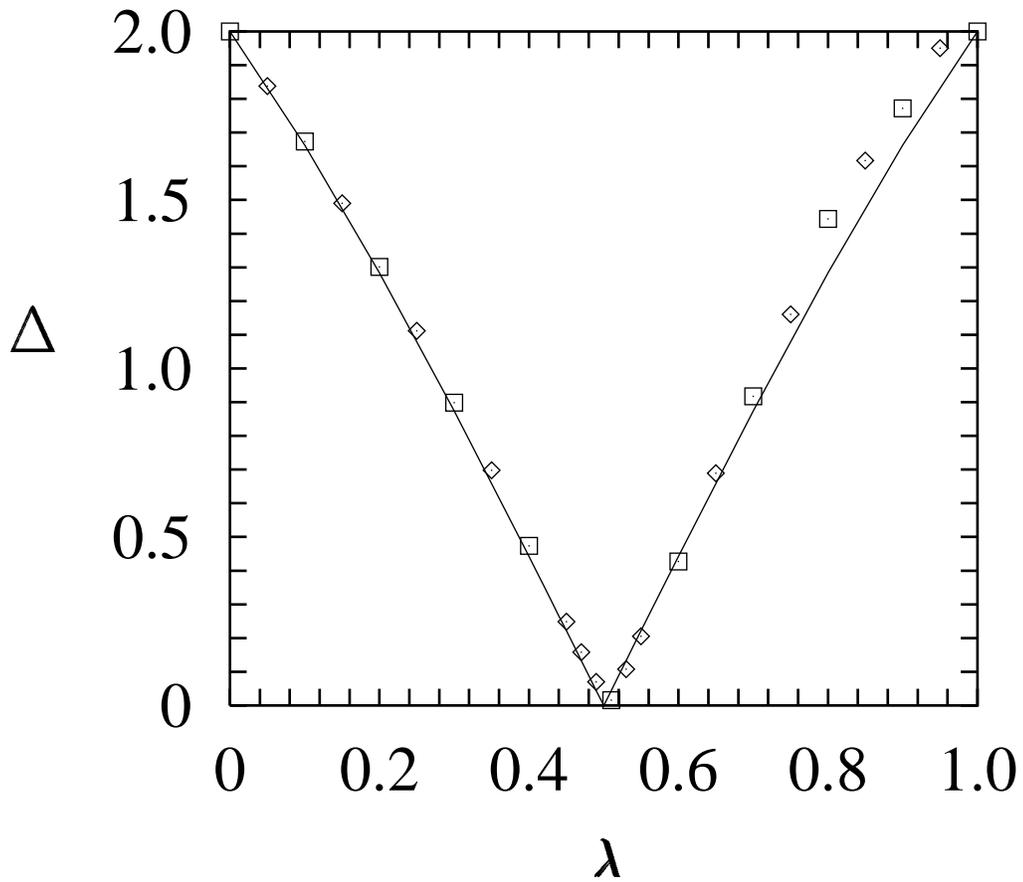}\end{center}
\caption[figgap]{Mass gap estimates $\Delta$ against $\lambda$.
The diamonds and squares indicate CORE estimates obtained using
$T_1^{16}$ and $T_2^{12}$, respectively.  The exact mass gap appears
as a solid curve.}
\label{figgap}
\end{figure}

\begin{figure}
\begin{center}\leavevmode\epsfbox[80 300 530 760]{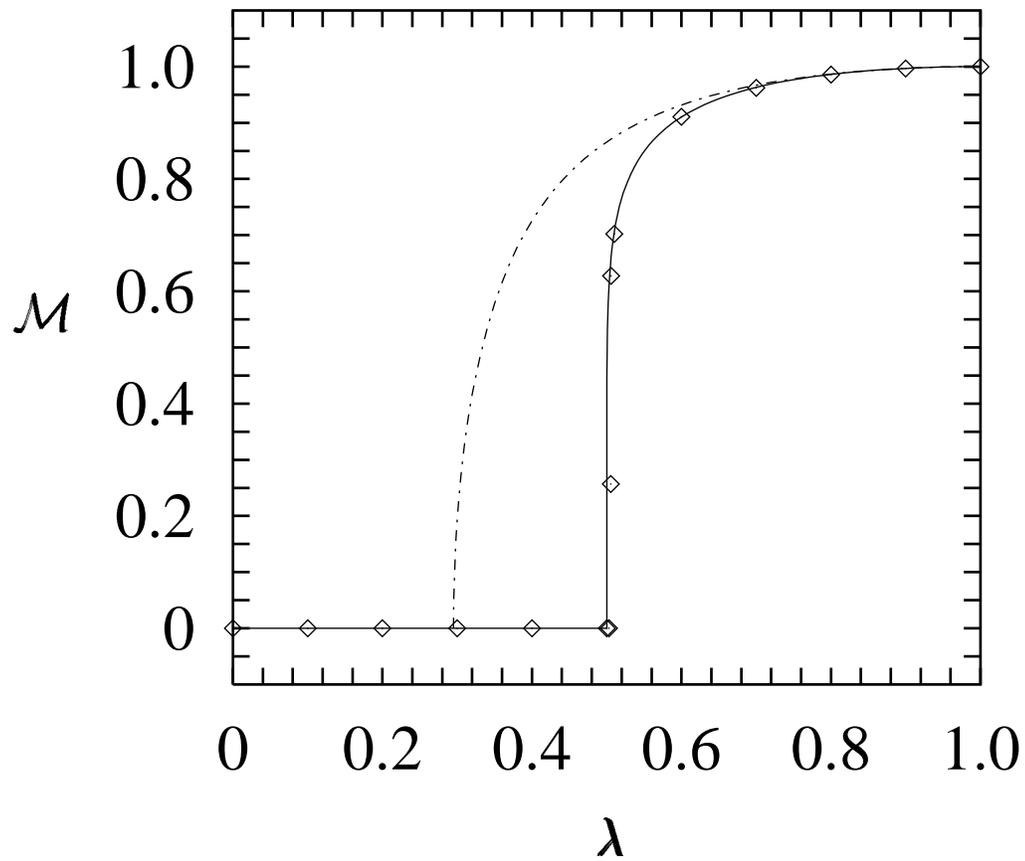}\end{center}
\caption[figmag]{Magnetization ${\cal M}$ against $\lambda$. The
diamonds indicate CORE estimates obtained using $T_2^{12}$,
the solid curve shows the exact magnetization, and the dot-dashed
curve shows the estimates from mean-field theory.}
\label{figmag}
\end{figure}

Selected estimates $E_0$ of the ground-state energy density
from both variants of the CORE approach described above
are compared to the exact \cite{isingexact} energy density $\epsilon_0$
in Fig.~\ref{figenergies}.  Calculations were done using
$T_1^n(t/n)$ and $T_2^n(t/n)$ for various values of $n$.
The fractional errors $\delta_{\epsilon_0}$
shown in this figure are defined by
$\delta_{\epsilon_0}=\vert (E_0-\epsilon_0)/\epsilon_0\vert$.
Selected mass gap estimates $\Delta$ are compared to the exactly-known
gap in Fig.~\ref{figgap}.
Fig.~\ref{figmag} illustrates the amounts by which the $T_2^{12}$
CORE estimates of the magnetization ${\cal M}=\vert\langle \sigma_x(j)
\rangle\vert$, for some site $j$, differ from the exact values.
The accuracy of the results is striking, especially considering
that only the first three terms in the cluster expansion were
included in the calculations.  The CORE method reproduces the
correct location of the critical point with remarkable precision.
The critical exponent $\zeta$ was extracted from a straight-line fit
of our $T_2^{12}$ results for $\ln{\cal M}$ to the form
$\ln {\cal M} = \zeta\,\ln(1-\Lambda_c^2/\Lambda^2)$,
where $\Lambda=\tan(\lambda\pi/2)$,
$\Lambda_c=\tan(\lambda_c\pi/2)$, and $\lambda_c$ is our
computed value for the critical point.  For $\lambda_c = 0.5053$ and
fitting in the range $0.51 \le \lambda \le 1.0 $, we obtain
$\zeta = 0.12437$, to be compared to the exact value of $0.125$.
The CORE procedure produces better results for a given effort than
multi-state RSRG methods previously used \cite{isingprev}.  The results
also compare very favorably to previous $t$-expansion calculations
\cite{texpan}.  Using larger blocks or including more terms in the cluster
expansion should further improve these results.

\paragraph*{Conclusion}
Given its simple theoretical foundations, the relative ease of
implementation, and our success in applying it to the 1+1-dimensional
Ising model, we believe that the CORE approximation will prove to be
a powerful tool for analyzing intrinsically nonperturbative systems.
One particularly exciting feature of this method is that it can be
applied to systems containing dynamical fermions, systems which
resist treatment by present stochastic means.  In general, we feel
that the possibility of eliminating the quenched approximation in lattice
quantum chromodynamics, better studying spontaneous symmetry breaking
and other nonperturbative phenomena in relativistic field theories,
and probing the low-energy physics of the Hubbard and $t-J$ models
warrants further work with the CORE approximation.

This work was supported by the U.~S.~DOE, Contract No.~DE-AC03-76SF00515,
NSERC of Canada, and the UK SERC, grant GR/J 21347.
\end{document}